\documentclass{raa_twocolumn}            

\usepackage{graphicx,amsmath,amssymb}             

\begin{document}

   \title{On the braking index evolution of PSR B0540-69: wind braking of pulsars
}

   \volnopage{Vol.0 (20xx) No.0, 000--000}      
   \setcounter{page}{1}          

   \author{H. Tong}

   \institute{School of Physics and Materials Science, Guangzhou University, Guangzhou 510006, China;
   {\it tonghao@gzhu.edu.cn}\\
   }

   \date{Received~~2009 month day; accepted~~2009~~month day}

\abstract{The pulsar PSR B0540-69 has both a braking index measurement and spin-down state change. After its spin-down state change, it shows an increasing braking index with time. Previously, it is pointed out that the spin-down state change  may be caused by an enhanced particle wind. The prediction is that its braking index  will be smaller in the high spin-down state. The current measured braking index is approaching the previous prediction. The transient variation of braking index may be due to a small varying part of the particle density. The  braking index evolution of PSR B0540-69 is found to be in an exponential form. Its braking index  is expected to approach some steady value. Future braking index measurement may make clear the physics for the braking index evolution. It may also help to make test the particle acceleration  mechanism in the pulsar magnetosphere. Finally, a phenomenological treatment of the wind braking model  is presented in the appendix. It can simplify the applications of wind braking model to pulsar braking index, intermittent pulsars and PSR B0540-69.
}

   \authorrunning{H.Tong}            
   \titlerunning{On the braking index evolution of PSR B0540-69}  

   \maketitle

%
%
\section{Introduction}
Timing and pulse profile are two observational aspects of pulsars. While pulse profile studies are related to the magnetospheric physics and radiation mechanism (Mignami et al. 2019). The timing of pulsars are mainly related to their internal and external dynamics (e.g., glitch and binary system, Lyne et al. 2022). Usually the period and period derivative of pulsars are reported during their discoveries (e.g., Han et al. 2021). For some pulsars, a meaningful frequency second derivative are also reported (Espinoza et al. 2017; Parthasarathy et al. 2020). The pulsar maybe slowing down in a power law form:
\begin{equation}
  \dot{\Omega} = -K \Omega^n,
\end{equation}
where $n$ is the so-called braking index, $\Omega$ and $\dot{\Omega}$ are the angular velocity and angular velocity time derivative of the neutron star, $K$ is some parameter depending on the detailed spin-down mechanism. For pure magnetic dipole braking, $n=3$. While for a neutron star spin-down by a particle wind (i.e. wind braking) $n\approx 1$ (Kou \& Tong 2015). Note that observationally the pulse frequency $\nu$ is often used. While theoretically, the angular velocity $\Omega$ or period $P$ are often cited. They describe the same rotational evolution of the neutron star. Assuming a constant parameter $K$, in observations the braking index is usually defined as:
\begin{equation}
  n= \frac{\Omega \ddot{\Omega}}{\dot{\Omega}^2}.
\end{equation}
Therefore, the pulsar timing observations, especially the braking index observations, can also be a powerful tool to diagnosis the braking mechanism and magnetospheric physics of pulsars (Xu \& Qiao 2001). Up to now, more than ten pulsars have braking index measurement (Espinoza et al. 2017; Parthasarathy et al. 2020).

Intermittent pulsars are the smoking gun evidence for the existence of a particle wind component in pulsar spin-down (Kramer et al. 2006; Camilo et al. 2012; Lorimer et al. 2012; Lyne et al. 2017). Taken the first intermittent pulsar as an example (Kramer et al. 2006), intermittent pulsars have on-state (with radio emission) and off-state (with no radio emission). Both the on-state and off-state can last for weeks or months. It is found that the intermittent pulsars have higher spin-down rate during the on-state than during the off state. The enhancement in spin-down rate can be $50\%$ to $150\%$ (Kramer et al. 2006; Camilo et al. 2012). A natural idea is that during the on-state there is a particle component in the pulsar spin-down torque. The appearance/disappearance of the particle component is responsible for the on/off state. Recent deep radio observations found weak radio emissions during the off-state (Rusul et al. 2026). This means that the off-state may have a weak particle component.

PSR B0540-69 has a period  of $P=50.7 \rm \ ms$. It has a high rotational energy loss rate about $1.5\times 10^{38} \rm\ erg \ s^{-1}$ (i.e. ``Crab twin"). Its characteristic age and characteristic magnetic field are $\tau_c =1.7\times 10^3 \rm yr$ and $B_c = 5\times 10^{12} \rm G$, respectively (Ho et al. 2026). PSR B0540-69 is special because it has both braking index measurement and a change in spin-down state. PSR B0450-69 has a braking index of $n=2.13$ during its persistent state (Ferdman et al. 2015). Later, it is found that its spin-down rate is enhanced by $36\%$ (Marshall et al. 2015). The change in spin-down rate is similar to that of intermittent pulsars. Therefore, both the braking index measurement and change in spin-down rate make PSR B0540-69 an excellent example to study pulsar braking mechanism (Kou et al. 2016, here after Paper I).

A general expectation is that during the high spin-down state, PSR B0540-69 will have a lower braking index. Quantitative calculations showed that it may have a new braking index about $n=1.79$ in the new spin-down state (Paper I, the detailed value depends on the particle acceleration potential in the wind braking model). Subsequent observations found that PSR B0540-69 stays in the new spin-down state and its braking is evolving with time. Marshall et al. (2016) found a braking of $n=0.031$ for PSR B0540-69. Later the braking index increase to about $n=1$ (Wang et al. 2020). Most recent measurements of braking index are about $n=1.4$ (Espinoza et al. 2024) and $n=1.6$ (Ho et al. 2026), respectively. The evolution of the braking index can be seen from the observational data in figure \ref{fig_gnobs}. It seems that the braking index is approaching the theoretical value $n=1.79$ predicted by the wind braking model.

Therefore, we think that the calculations in Paper I are still valid for the new spin-down state. And it can be tested by further braking index observations of PSR B0540-69. The calculations will not be repeated here. This time, we focus on the transient evolution of the braking index. If the braking index during the persistent state and change in spin-down rate are due to the particle component. Then a transient evolving component in the particle density may account for the evolution of the braking index. That is: a small and transient evolving component in the particle density will not affect the spin-down rate $\dot{\Omega}$ (PSR B0540-69 still in the high spin-down state), but it will affect the braking index measurement (which is related to $\ddot{\Omega}$). This is the main idea of this paper.

\section{Wind braking model for the braking index evolution of PSR B0540-69}

\subsection{Summary of the wind braking model}

The details of the wind braking model can be found in Kou \& Tong (2015), Tong \& Kou (2017). A phenomenological treatment of the wind braking model in presented in the Appendix. In the wind braking model, the rotational energy of the neutron star is carried away by both the magnetic dipole radiation and the outflow of a particle wind. Quantitatively, it is:
\begin{equation}\label{eqn_rotational_energy_loss}
  -I \Omega \dot{\Omega} = \frac{2\mu^2 \Omega^4}{3c^3} \eta,
\end{equation}
where $I$ is the moment of inertia of the neutron star, $\mu$ is the magnetic dipole moment, and $\eta$ is the dimensionless parameter characterizing the wind braking model. For an inclination angle $\alpha$, a particle number density $\kappa$ times the Goldreich-Julian density, and an acceleration potential $\Delta \phi$, the dimensionless parameter is:
\begin{equation}
  \eta = \sin^2 \alpha + 3 \kappa \frac{\Delta \phi}{\Delta \Phi} \cos^2\alpha,
\end{equation}
where $\Delta \Phi$ is the maximum acceleration potential of the rotating magnetized neutron star. The first term in $\eta$ (proportional to $\sin^2\alpha$ ) is the magnetic dipole radiation term. The second term in $\eta$ is the particle wind term. In the absence of a particle wind, the wind braking model reduces to the magnetic dipole braking case. In the second term of $\eta$, the dimensionless particle density always appear in combination with inclination angle, i.e., in the form $\kappa \cos^2\alpha$. Therefore, if the second term is without a factor of $\cos^2\alpha$, the particle number density $\kappa$ only changes quantitatively. The acceleration potential is expressed in units of the maximum acceleration potential. For a given acceleration potential, the dimensionless parameter $\eta$ can be determined (see Kou \& Tong 2015). The vacuum gap model (Ruderman \& Sutherland 1975) is often taken as an example. In the vacuum gap model, the dimensionless $\eta$ parameter is:
\begin{equation}\label{eqn_eta_VG}
  \eta = \sin^2 \alpha + 4.96 \times 10^2 \kappa B_{12}^{-8/7} \Omega^{-15/7} \cos^2\alpha.
\end{equation}
The rotational evolution of the pulsar in the wind braking model can be denoted as:
\begin{equation}\label{eqn_dotOmega_wind}
  \dot{\Omega} = -\frac{2\mu^2 \Omega^3}{3 I c^3} \eta \equiv - K \Omega^3 \eta.
\end{equation}
From this equation and the definition of braking index, it can be seen that the magnetic dipole radiation term has a braking index of $n=3$. While the particle wind term has a braking index of $n=6/7 \approx 1$.

\subsection{On the braking index evolution of PSR B0540-69}

\begin{figure}
  \centering
  \includegraphics[width=0.5\textwidth]{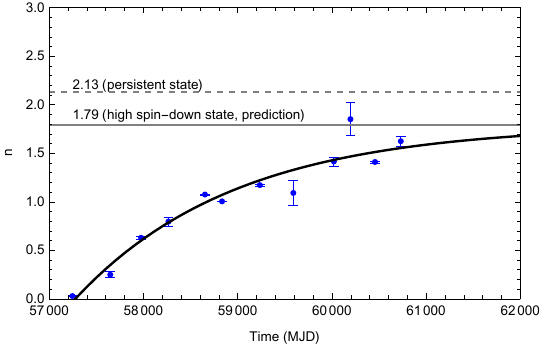}\\
  \caption{Braking index evolution of PSR B0540-69 with time. The blue points are the observational data from Table 6 in Ho et al. (2026). The solid line is the model calculations in the wind braking model, eq.(\ref{eqn_n_time_evo}). The gray dashed line is the braking index of PSR B0540-69 in the persistent state (Ferdman et al. 2015). The gray solid line is the prediction of Kou et al. (2016) for the high spin-down state. }\label{fig_gnobs}
\end{figure}

For a constant particle density, taking the time derivative of eq.(\ref{eqn_dotOmega_wind}), the braking index in the wind braking model is (Kou \& Tong 2015):
\begin{equation}\label{eqn_braking_index_persistent}
  n = 3+ \frac{\Omega}{\eta} \frac{d\eta}{d\Omega}.
\end{equation}
This may account for the braking index of the Crab pulsar, PSR B0540-69, and other pulsars with braking index measurement.

The particle density has a value of $\kappa_0$ in the persistent state. Due to some changes in the magnetosphere, the particle density increase from $\kappa_0$ to $\kappa_1$. Then the pulsar will have a higher spin-down rate. And in the high spin-down state, the pulsar is expected to have a lower braking index. This may correspond to the spin-down state change in PSR B0540-69 and the detailed calculations are given in Paper I.

It is possible that when PSR B0540-69 changes to the new spin-down state, the particle density may have a small  transient variation part. That is:
\begin{equation}\label{eqn_kappa_decay}
  \kappa = \kappa_1 (1 + \delta(t) ) = \kappa_1 (1 + \delta e^{-t/\tau}),
\end{equation}
where $\delta(t) = \delta e^{-t/\tau}$ (an exponential decaying form is assumed), $\delta$ is the amplitude of $\delta(t)$, $\tau$ is the decaying timescale. It is possible that $|\delta| \ll 1$. Then the transient varying part of the particle density will not affect the spin-down rate. But it may contribute to a transient varying $\ddot{\Omega}$, i.e. braking index. The $\ddot{\Omega}$ contributed by the transient varying part of the particle density is:
\begin{eqnarray}
  \ddot{\Omega} &=& - K \Omega^3 \frac{d\eta}{d\kappa} \dot{\kappa} \\
  &=& \dot{\Omega} \frac{1}{\eta} \frac{d\eta}{d\kappa} (-\frac{\kappa_1}{\tau}) \delta e^{-t/\tau}.
\end{eqnarray}
The modification to the braking index (in addition to that of eq.(\ref{eqn_braking_index_persistent})) is:
\begin{equation}
  \Delta n(t)= 2 \frac{\kappa_1}{\eta} \frac{d\eta}{d\kappa} \frac{\tau_c}{\tau} \delta e^{-t/\tau},
\end{equation}
where $\tau_c \equiv P/(2\dot{P}) = -\Omega/(2\dot{\Omega})$ is the characteristic age of the pulsar. Irrespective of the detailed form of $\eta$, the modification to the braking index is about:
$(\tau_c/\tau) \delta$. Although $|\delta | \ll 1$, but for $\tau_c/\tau \gg 1$, the transient varying part of the particle density may affect the braking index significantly. The requirement is that $(\tau_c/\tau) \delta \sim O(1)$. And the modification to the braking index is decaying with time, e.g. in an exponential form. The modification to the braking index can be combined in a form:
\begin{equation}
  \Delta n(t) = \Delta n e^{-t/\tau},
\end{equation}
where $\Delta n$ is the amplitude of the varying part of the braking index $\Delta n(t)$. Combined with the braking index in eq.(\ref{eqn_braking_index_persistent}), the total braking of PSR B0540-69 may be written in the form:
\begin{equation}\label{eqn_n_time_evo}
  n(t) = n_{\rm new} + \Delta n(t) = 1.79 + \Delta n e^{-t/\tau},
\end{equation}
where the braking index in the high spin-down state is chosen as $n_{\rm new} = 1.79$, the result of the vacuum gap case (Paper I). Irrespective of the detailed modelings (see discussions in the Appendix), the braking index is expected to be in an exponential decaying form. This is the consequence of an exponential decaying variation of the particle density (eq.(\ref{eqn_kappa_decay})).

The theoretical curve is shown in figure \ref{fig_gnobs}. For the theoretical curve, the typical parameters are: $\Delta n =-2.1$, $\tau= 1708$ days. Therefore, the varying amplitude of the particle density is about: $\delta \sim \tau/\tau_c \sim 10^{-3}$. Consistent with our starting point that it will not affect the spin-down rate, and will mainly contribute to the transient varying braking index. During the calculations, the start of the time is chosen as  MJD 57000, near  the time of the first braking index measurement of PSR B0540-69 (first line in Table 6 of Ho et al. 2026).

From  figure \ref{fig_gnobs}, the theoretical curve can catch the general trend of the braking index evolution of PSR B0540-69. And it has clear predictions: the transient varying braking index of PSR B0540-69 will settle to a constant value in the near future. The final value will be the braking index in the high spin-down state of PSR B0540-69. In the wind braking model it should be smaller than the persistent state braking (e.g., $n=1.79$ in the vacuum gap case). Therefore, the braking index of PSR B0540-69 in the near future can constrain the particle acceleration unambiguously. At that time, PSR B0540-69 will have two spin-down states and two braking index measurements. This will make it an excellent example to study pulsar braking mechanism.

\section{Discussion and  conclusion}

There are two predictions here. (1) The braking index of PSR B0540-69 is expected to evolve in an exponential form (second term in eq.(\ref{eqn_n_time_evo})). This is demonstrated in the main text under the wind braking model . This point may also be caused by exponential variations of the magnetic field, moment of inertial, or inclination angle (see the appendix for discussions). It is independent of the model details. (2) The braking index of PSR B0540-69 will approach some steady value in the near future (first term in eq.(\ref{eqn_n_time_evo})). In the wind braking model, the steady value of the braking index will approach that predicted in Paper I. This point is only valid in the wind braking scenario.

Since Marshall et al. (2016), it is evident that the braking index of PSR B0540-69 is evolving with time. There are several proposals for the braking index evolution of PSR B0540-69. These include: (1) evolution of magnetic field ($\dot{B}$, Wang et al. 2020), (2) evolving moment of inertial ($\dot{I}$, Rusul et al. 2023), (3) evolving inclination angle ($\dot{\alpha}$, Barao et al. 2026). These aspects are also employed when explaining the braking index of other pulsars (Lyne et al. 2015).  By looking at the braking index evolution of PSR B0540-69 in the high spin-down state alone, it may be hard to judge which is the correct explanation. However, the existence of intermittent pulsars tells us clearly that the particle density is responsible for the enhanced spin-down during the on-state (Kramer et al. 2006). Therefore, the spin-down state change of PSR B0540-69 may also be caused by an enhanced particle density (Paper I). Then it is natural that there is some small transient variation of the particle density.

Furthermore, for an evolving magnetic field $\dot{B}$ (or moment of inertia, or inclination angle, which  is similar), a finite $\dot{B}$ may explain the persistent state braking index of PSR B0540-69 ($n=2.13$, Ferdman et al. 2015). A discrete change in the magnetic field strength may explain the spin-down state change (Marshall et al. 2015). However, in the new high spin-down state, the changing magnetic field model will lose its prediction power. That is: the final braking index in the high spin-down state can be an arbitrary value. A natural expectation is that the braking index in the high spin-down state will return to that of the persistent state (Wang et al. 2020). Since the recent ten years is a short period in the life time of PSR B0540-69. It will try to go back to its previous persistent state. In contrast, the wind braking model has clear predictions: PSR B0540-69 will have a smaller braking index during the high spin-down state ($n=1.79$ in the vacuum gap case). The exact value depends on the particle acceleration potential.

Recent observations of intermittent pulsars found weak emission during the off-state (Rusul et al. 2026). This means that there may be a weak particle component in the ``off-state".  This will make  intermittent pulsars and PSR B0540-69 more akin to each other. A phenomenological treatment of the wind braking model is presented in the Appendix. More discusses related to the pulsar braking index, intermittent pulsars, and PSR B0540-69 are given there.

In conclusion, the high spin-down state of PSR B0540-69 may be caused by an enhanced particle wind (Paper I). The particle wind may have a small transient variation. This may explain the braking index evolution of PSR B0540-69. The braking index of PSR B0540-69 is expected to evolve in an exponential form (eq.\ref{eqn_n_time_evo}). Future braking index observations of PSR B0540-69 will help to constrain the underlying physics and the particle acceleration mechanism in the pulsar magnetosphere.

\appendix

\section{A phenomenological treatment of the wind braking model of pulsars}

As been pointed out in the text, magnetic dipole braking corresponds to a braking of $n=3$. While, the particle wind term has braking index about $n\approx 1$. Noting this point, some general information can be obtained for the wind braking model (Lyne et al. 2015; Kou \& Tong 2015 for previous discussions). In the following, we present a more detailed treatment. This phenomenological treatment can be view as a simplified version of the wind braking model.

In general, the wind braking model contains a dipole term and a particle wind term. Therefore, the rotational evolution of the pulsar in the wind braking model is:
\begin{equation}\label{eqn_Omegadot_phenomenological}
  \dot{\Omega} = -K_1 \Omega^3 -K_2 \Omega,
\end{equation}
where $K_1$ and $K_2$ are numerical constants depending on the detailed modeling. From the above equation, in the spin-down torque, the dipole term contribution is: $-K_1 \Omega^3$, the particle wind term is: $-K_2 \Omega$. These two terms may be different in different sources. The fraction of particle wind in the total torque may be more important:
\begin{equation}
  f_{\rm w} = \frac{-K_2 \Omega}{-K_1 \Omega^3 -K_2 \Omega} = \frac{K_2 \Omega}{K_1 \Omega^3 +K_2 \Omega}.
\end{equation}
This phenomenological treatment can be applied to (1) the braking index of pulsars, (2) intermittent pulsars, and (3) PSR B0540-69.

(1) \textit{Braking index of pulsars.} From the expression of $\eta$ in the wind braking model, the above parameters $K_1$ may depend on the star's magnetic field, moment of inertia, and inclination angle: $K_1=K_1(B, I, \sin\alpha)$. While $K_2$ may depend on the star's magnetic field, moment of inertia, particle number density, and inclination angle: $K_2 =K_2(B, I, \kappa, \cos\alpha)$. For a constant $K_1$ and $K_2$, from the definition of braking index, the braking index in the phenomenological treatment is:
\begin{equation}
  n = 3-2 f_{\rm w}.
\end{equation}
This equation tells us that the deviation of the pulsar's braking index from $n=3$ is determined by the fraction of particle wind $f_{\rm w}$. Furthermore, for a pulsar with $\dot{\Omega}$ and braking index $n$ measurement, we will have two equations for the two parameter $K_1$ and $K_2$. Their exact value can be determined.

(2) \textit{Spin-down ratio of intermittent pulsars.} During the on-state, the intermittent pulsar is spin-down by both the dipole radiation and particle wind. While, in the off-state the particle wind term may be absent. Therefore, its spin-down rate during the on-state and off-state are:
\begin{eqnarray}
  \dot{\Omega}_{\rm on} &=& -K_1 \Omega^3 -K_2 \Omega, \\
  \dot{\Omega}_{\rm off} &=& -K_1 \Omega^3.
\end{eqnarray}
Therefore, the ratio of spin-down rate for intermittent pulsars is:
\begin{equation}
  R_{\dot{\Omega}} = \frac{\dot{\Omega}_{\rm on}}{\dot{\Omega}_{\rm off}} = \frac{-K_1 \Omega^3 -K_2 \Omega}{-K_1 \Omega^3} = \frac{1}{1-f_{\rm w}}.
\end{equation}
Therefore, the spin-down ratio for intermittent pulsars is also related to the particle fraction parameter $f_{\rm w}$. Combining with the braking index relation, the predicted braking index for intermittent pulsars during their on state is:
\begin{equation}\label{eqn_n_intermittent_pulsar}
  n_{\rm on} = 1+ \frac{2}{R_{\dot{\Omega}}}.
\end{equation}
This may be tested by future observations of intermittent pulsars.

(3) \textit{PSR B0540-69.} During the persistent state of PSR B0540-69, its spin-down rate can be denoted as:
\begin{equation}
  \dot{\Omega}_{\rm persistent} = -K_1 \Omega^3 - K_2 \Omega.
\end{equation}
A higher particle wind may correspond to the high spin-down state of PSR B0540-69. The spin-down rate in the high spin-down state is:
\begin{equation}
  \dot{\Omega}_{\rm new} = -K_1 \Omega^3 -K_2^\prime \Omega,
\end{equation}
where $K_2^\prime$ corresponds to the new particle wind term. The increase of the particle wind is responsible for the new spin-down rate. Denote the enhancement of particle wind as:
\begin{equation}
  e_{\rm w} = \frac{-K_2^\prime \Omega}{-K_2 \Omega} = \frac{K_2^\prime}{K_2}.
\end{equation}
Then the spin-down ratio for PSR B0540-69 is:
\begin{equation}
  R_{\dot{\Omega}} = \frac{\dot{\Omega}_{\rm new}}{\dot{\Omega}_{\rm persistent}} = 1+ (e_{\rm w}-1) f_{\rm w}.
\end{equation}
Therefore, from the observed spin-down ratio $R_{\dot{\Omega}}$, the enhancement in the particle wind $e_{\rm w}$ can be determined.
Once the enhancement parameter $e_{\rm w}$ is determined, the corresponding particle wind fraction in the new spin-down state can also be obtained. Then, the predicted braking index for PSR B0540-69 in the new spin-down state is:
\begin{equation}\label{eqn_n_B0540_phenomonological}
  n_{\rm new} = 1+ \frac{n-1}{R_{\dot{\Omega}}},
\end{equation}
where $n$ is the braking index in the persistent state. Equation (\ref{eqn_n_B0540_phenomonological}) and (\ref{eqn_n_intermittent_pulsar}) are similar to each other.

Quantitatively, PSR B0540-69 has a braking index of $n=2.13$ during the persistent state (Ferdman et al. 2015). Then the particle wind fraction in its spin-down torque is: $f_{\rm w} = 0.435$. For a $36\%$ increase in spin-down rate (i.e. $R_{\dot{\Omega}} = 1.36$, Marshall et al. 2015), the enhancement in its particle wind is about $83\%$ (i.e. $e_{\rm w} = 1.83$) (compared with $88\%$ increase in the vacuum gap case, see Paper I). The predicted braking index for PSR B0540-69 in the high spin-down state is: $n_{\rm new} = 1.83$ (compared with 1.79 in the vacuum gap case, see Paper I).

The above treatment (eq.(\ref{eqn_Omegadot_phenomenological})) can be generalized to include a particle wind term with a general braking index, e.g. $\dot{\Omega} = -K_1 \Omega^3 -K_2 \Omega^{n_{\rm wind}}$, where the braking index of the particle wind term is $n=n_{\rm wind}$. As a pulsar slows down (i.e. $\Omega$ decreases), it will evolve from dipole braking dominated to that of wind braking dominated. Its braking index will evolve from $n=3$ to $n=1$. However, for the long term evolution of pulsars, the above phenomenological treatment will miss some important physics: (1) the particle wind will diminish when the pulsar period approaches the death line. (2) The rotational evolution of the pulsar is coupled with the inclination angle evolution. These two  aspects need physical treatment of the wind braking model (Kou \& Tong 2015; Tong \& Kou 2017).

If the particle wind term parameter $K_2$ has some small variation during the persistent state (i.e. $K_2 \rightarrow K_2 (1+ \delta(t))$), then it may corresponds to the variation in spin-down torque ($\dot{\Omega}$), or fluctuations in the $\ddot{\Omega}$, or timing noise (Hobbs et al. 2010; Ou et al. 2016; Lower et al. 2025). Similar things may also happen for the high spin-down state. If during the high spin-down state, the particle wind term has some small transient variations (i.e. $K_2^\prime \rightarrow K_2^\prime(1+ \delta(t))$), this may result in the braking index evolution of PSR B0540-69, as been discussed in the main text. Furthermore, the dependence of $K_2=K_2(B, I, \kappa, \cos\alpha)$ on the magnetic field, moment of inertial, particle density, and inclination angle are all in a power law form. Then either variation of the four parameters $(B, I, \kappa, \cos\alpha)$ will result in similar variation of $K_2$: $K_2^\prime \rightarrow K_2^\prime(1+ \delta(t))$. However, from the combined information of intermittent pulsar observations, pulsar braking index measurements, and pulsar magnetosphere simulations, the variation in particle density may be more possible.

\section*{acknowledgments}
This work is supported by NSFC (12133004) and National SKA Program of China (No.2020SKA0120300).




\label{lastpage}

\end{document}